\documentclass[11pt, onecolumn]{article}
\usepackage[top=1in, bottom=1in, left=1.25in, right=1.25in]{geometry}

\usepackage{amssymb}
\usepackage{amssymb}
\usepackage{amsmath}
\usepackage{amssymb}
\usepackage[amsmath,thmmarks]{ntheorem}
\usepackage{indentfirst}

\usepackage[dvips]{graphicx}

\begin{document}

\title{Retrieval of Sparse Solutions of Multiple-Measurement Vectors via Zero-point Attracting Projection}

\author{Yang~You,~Laming~Chen,~Yuantao~Gu\thanks{This work was partially supported by National Natural Science Foundation of China (NSFC 60872087 and NSFC U0835003). The corresponding author of this paper is Yuantao Gu (Email: gyt@tsinghua.edu.cn).},~Wei~Feng,~Hui~Dai}

\date{Received Nov. 2, 2011; accepted June 7, 2012.\\\vspace{1em}
This article appears in \textsl{Signal Processing}, 92(12): 3075-3079, 2012.}

\maketitle

\begin{abstract}
A new sparse signal recovery algorithm for multiple-measurement vectors (MMV) problem is proposed
in this paper. The sparse representation is iteratively drawn based on the idea of zero-point
attracting projection (ZAP). In each iteration, the solution is first updated along the negative
gradient direction of an approximate $\ell_{2,0}$ norm to encourage sparsity, and then projected to
the solution space to satisfy the under-determined equation. A variable step size scheme is adopted
further to accelerate the convergence as well as to improve the recovery accuracy. Numerical
simulations demonstrate that the performance of the proposed algorithm exceeds the references in
various aspects, as well as when applied to the Modulated Wideband Converter, where recovering MMV
problem is crucial to its performance.

\textbf{Keywords:} compressed sensing, sparse recovery, multiple-measurement vectors, zero-point attracting projection, the Modulated Wideband Converter.
\end{abstract}

\section{Introduction}
\label{sec:intro}

Sparse recovery is one of the essential issues in many fields of signal processing, including
compressed sampling (CS) \cite{Candes, Donoho}, which is a novel sampling theory. In some
applications such as magnetoencephalography (MEG) \cite{Cotter}, source localization
\cite{Malioutov} and analog-to-digital conversion \cite{Eldar}, the sparse unknowns can be
recovered jointly, which results in the problem of multiple-measurement vectors (MMV). In the
Modulated Wideband Converter (MWC) \cite{Eldar}, the locations of narrow-band signals in a wide
spectrum range are detected by solving the MMV problem, which is of great significance to the
performance of the conversion system.

Several algorithms have been proposed to derive the sparse solution to the MMV problem. The OMP
algorithm for MMV \cite{JieChen} finds a sparse solution by sequentially building up a small subset
of column vectors selected from ${\bf A}$ to represent ${\bf Y}$. FOCal Underdetermined System
Solver (FOCUSS) methods \cite{Cotter} compute the constraint solution in an iterative procedure
using the standard method of Lagrange multipliers. ReMBo \cite{Mishali} reduces MMV to SMV by
multiplying a random column vector drawn from an absolutely continuous distribution. Recently, a
robust algorithm for joint-sparse recovery named Joint $\ell_{2,0}$ Approximation algorithm (JLZA)
\cite{Hyder} has been proposed to calculate the solution in a fixed point iteration with an
approximation of $\ell_{2,0}$ norm. Different algorithms which aim at solving the problem of MMV
are surveyed and compared in \cite{Rakotomamonjy}.

This work extends a recently proposed zero-point attracting projection (ZAP) algorithm \cite{Jin}
to the MMV scenario. Derived from the adaptive filtering framework, ZAP defines the cost function
as the sparsity and updates the sparse representation in the solution space. Specifically, starting
from the least squares solution, it is modified in a fixed step-size along the negative gradient
direction of an approximate $\ell_0$ norm towards the zero point, and then projected back to the
solution space again. This operation is executed iteratively until the stopping condition is
satisfied. In this paper, an approximate $\ell_{2,0}$ norm is adopted in ZAP and the recursive
solution to MMV is derived. In addition, the step-size of zero-point attraction is varied
dynamically to accelerate the convergence. The rest of the present paper is organized as follows.
Section~\ref{sec:prob} formulates the MMV problem and the approximate $\ell_{2,0}$ norm.
Section~\ref{sec:ZAPMMV} introduces the ZAP algorithm for MMV, and further discussions and
improvements are given in Section~\ref{sec:dis}. Section~\ref{sec:simu} shows the simulation
results and this paper is concluded in Section~\ref{sec:conclu}.

\section{Problem Formulation}
\label{sec:prob}

The problem of MMV is formulated as ${\bf Y}={\bf A}{\bf X}$, where ${\bf Y}\in\mathbb{R}^{M\times
L}$ is the matrix of measurements, ${\bf A}\in\mathbb{R}^{M\times N}$ is the sensing matrix, and
${\bf X}\in \mathbb{R}^{N\times L}$ is the unknown signal assumed jointly $K$-sparse, i.e. it has
$K$ nonzero rows at most. It is shown \cite{JieChen} that the following problem
\begin{align}\label{mmv}
\hat{\bf X} = \arg\min_{\bf X}\|{\bf r}({\bf X})\|_0, \ {\rm subject\ to}\ {\bf Y}={\bf A}{\bf X}
\end{align}
finds the unique $K$-sparse solution provided that
\begin{align*}
K<\frac{{\rm Spark}({\bf A})+{\rm Rank}({\bf Y})-1}{2},
\end{align*}
where ${\bf r}({\bf X})$ returns a column vector whose $i$th item is any vector norm of the $i$th
row of ${\bf X}$, $\|\cdot\|_0$ counts nonzero elements of the vector, ${\rm Spark}({\bf A})$ is
the smallest possible integer such that there exist ${\rm Spark}({\bf A})$ columns of matrix $\bf
A$ that are linearly dependent, and ${\rm Rank}({\bf Y})$ is the rank of matrix $\bf Y$.

According to the references \cite{Cotter, Hyder}, we choose $\ell_{2,0}$ norm
\begin{align}\label{l20norm}
J({\bf X})=\|{\bf r}_2({\bf X})\|_0
\end{align}
to describe the joint sparsity of $\bf X$, where ${\bf r}_2({\bf X})$ denotes a column vector with
$\ell_2$ norm of the $i$th row of ${\bf X}$ as its $i$th item. Following \cite{Weston,Jin}, the
sparsifying penalty can be further approximated by a continuous function
\begin{align}\label{appl20norm}
J({\bf X})=\sum_{i=1}^N F_{\alpha}(\|{\bf x}_i\|_2),
\end{align}
where ${\bf x}_i$ denotes the $i$th row of ${\bf X}$, and the function $F_{\alpha}(\cdot)$ is
defined as
\begin{align}\label{Falpha}
F_{\alpha}(w)=\left\{
\begin{array}{cl} 2\alpha |w|-{\alpha^2}w^2 & |w|\leq\frac{\textstyle1}{\textstyle\alpha}; \\
1 & {\rm elsewhere}. \end{array}\right .
\end{align}
It is readily recognized that $F_{\alpha}(w)$ converges to the $\ell_0$ quasi-norm as $\alpha>0$
approaches infinity. Since it is non-differentiable at the origin, the derivative of
$F_{\alpha}(\cdot)$ can be replaced by one of its sub-derivatives
\begin{align}
f_{\alpha}(w)=\left\{
\begin{array}{cl} 2\alpha\cdot {\rm sgn}(w)-2{\alpha^2}w & |w|\leq\frac{1}{\textstyle\alpha}; \\
0 & {\rm elsewhere}, \end{array} \right. \label{falpha}
\end{align}
where sgn($\cdot$) denotes the sign function with sgn$(0)=0$.

\section{Zero-point Attracting Projection for MMV}
\label{sec:ZAPMMV}

In this section we extend ZAP to the MMV scenario. Further discussions and improvements of this
algorithm will be shown in section \ref{sec:dis}.

ZAP for MMV derives the sparse solution in an iterative procedure. In the $n$th iteration, the
solution is first updated along the negative gradient direction of the sparsifying penalty
(\ref{appl20norm}),
\begin{align}\label{update}
\tilde{\bf X}(n)={\bf X}(n-1)-\kappa\cdot\nabla J({\bf X}(n-1)),
\end{align}
where $\tilde{\bf X}(n)$ denotes the temporary solution, $\kappa>0$ denotes the step size, and
$\nabla J(\cdot)$ is a gradient matrix with partial derivative
\begin{align}\label{partial}
\left(\nabla J({\bf X})\right)_{i,j} = \frac{\partial J({\bf X})}{\partial x_{i,j}}
\end{align}
as its $(i,j)$th element, where $x_{i,j}$ stands for the $(i,j)$th entry of ${\bf X}$. Substituting
(\ref{Falpha}) into (\ref{appl20norm}), and together with (\ref{falpha}), the $(i,j)$th element of
the gradient matrix can be calculated
\begin{align}\label{partial1}
\left(\nabla J({\bf X})\right)_{i,j}=\frac{\partial F_{\alpha}(\|{\bf x}_i\|_2)}{\partial x_{i,j}}
=\left\{
\begin{array}{cl}
\frac{\textstyle f_{\alpha}(\|{\bf x}_i\|_2)}{\textstyle \|{\bf x}_i\|_2}\cdot x_{i,j} & \|{\bf x}_i\|_2\neq 0;\\
0 & \|{\bf x}_i\|_2=0.\end{array} \right.
\end{align}
Then the solution is projected to the solution space to satisfy $\bf Y=AX$
\begin{align}\label{projection}
{\bf X}(n)=\tilde{\bf X}(n)+{\bf A}^{\dagger}({\bf Y}-{\bf A}\tilde{\bf X}(n)),
\end{align}
where ${\bf A}^{\dagger}={\bf A}^{\rm T}({\bf A}{\bf A}^{\rm T})^{-1}$ denotes the pseudo-inverse
matrix of ${\bf A}$.

Now we have introduced the basic procedure of ZAP for MMV. Compared with solving the MMV problem
column by column independently, ZAP for MMV utilizes the information of joint sparsity by means of
(\ref{partial1}). The updating step size of each element is renormalized according to the norm of
the corresponding row, which greatly improves the recovery performance. Detailed discussions about
the effect of the parameters and a fast convergence version of the proposed algorithm are given in
the following section.

\section{Discussions and Improvements}
\label{sec:dis}

So far, the parameters involved in the ZAP for MMV are the $\ell_0$ norm approximation factor
$\alpha$ and the step size $\kappa$. These parameters should be selected carefully.

{\bf{\emph{The choice of $\alpha$}}}: The value of $\alpha$ has a great impact on the performance
of the proposed algorithm. By (\ref{falpha}) we know that large $\alpha$ pulls small values to zero
quickly but the range of its influence is small, while a small one has the opposite effect.
Furthermore, according to (\ref{Falpha}), larger $\alpha$ leads to a better approximation of the
$\ell_0$ norm, but simultaneously produces more local minima in the cost function \cite{Hyder}. In
practice, we suggest $1/\alpha$ to be a quarter to a half of the smallest nonzero item of ${\bf
r}_2({\bf X})$.

{\bf{\emph{The choice of $\kappa$}}}: As the step size in the gradient descent iterations, the
parameter $\kappa$ suggests a tradeoff between the speed of convergence and the accuracy of the
solution. Big $\kappa$ indicates faster convergence but less accuracy, while small $\kappa$ yields
more accurate solution but costs more iterations. In order to improve the performance of the
proposed algorithm, the idea of variable step size is taken into consideration. The control scheme
is rather direct: $\kappa$ is initialized to be a large value, and reduced by a factor $\eta$ when
the algorithm is convergent. This reduction is conducted until $\kappa$ is sufficiently small, i.e.
$\kappa<\kappa_{\rm min}$, which means that the recovery error reaches a low level.

{\bf{\emph{The criterion of convergence}}}: It is significantly important to select an appropriate
criterion of convergence to make full use of variable step size. Considering the iteration is
performed along the negative gradient direction of the sparsifying penalty (\ref{appl20norm}), it
is obvious that the algorithm reaches steady-state when the penalty starts increasing. To reduce
the complexity of evaluation and improve the stability, the criterion is checked every $Q$
iterations.

{\bf{\emph{The initial value}}}: Following the ZAP algorithm for SMV, the least squares solution is
selected as the initial value, i.e. ${\bf X}(0) = {\bf A}^{\dagger}{\bf Y}$.

{\bf{\emph{The stopping condition}}}: The proposed algorithm stops (a) when $\kappa$ gets below
$\kappa_{\rm min}$ or (b) when the number of iterations reaches the bound $T$.

Finally, the proposed algorithm denoted by ZAPMMV is described in Table~\ref{codeofalgorithm}.

\begin{table}[t]
\caption{The ZAPMMV Algorithm}
\begin{center}\label{codeofalgorithm}
\begin{tabular}{l}
\hline {\bf Initialization}\\ \hspace{1.5em}1. Calculate ${\bf A}^{\dagger}={\bf A}^{\rm
T}({\bf
A}{\bf A}^{\rm T})^{-1}$; \\
\hspace{1.5em}2. Set ${\bf X}(0)={\bf A}^{\dagger}{\bf Y}$;\\
\hspace{1.5em}3. Set $\alpha$, $\kappa$, $\eta$, $\kappa_{\rm min}$, $Q$, $T$;\\
\hspace{1.5em}4. $n=1$;\\
{\bf repeat}\\
\hspace{1.5em}5. Calculate $\nabla J({\bf X}(n-1))$ by (\ref{falpha}) and (\ref{partial1});\\
\hspace{1.5em}6. Update ${\bf X}(n)$ by (\ref{update});\\
\hspace{1.5em}7. Project ${\bf X}(n)$ to the solution space by (\ref{projection});\\
\hspace{1.5em}8. If mod$(n,Q)=0$ and $J({\bf X}(n))\ge J({\bf X}(n-Q))$\\
\hspace{4em}$\kappa=\eta\kappa$;\\
\hspace{2.6em}End if \\
\hspace{1.5em}9. Iteration number increases by one: $n=n+1$;\\
{\bf until} $\kappa<\kappa_{\rm min}$ or $n=T$.\\
\hline
\end{tabular}
\end{center}
\end{table}

\section{Simulations}
\label{sec:simu}

In this section four experiments are designed to demonstrate the performance of the proposed
algorithm in various aspects, including recovery probability versus sparsity and robustness to
measurement noise, the running time of the algorithms, as well as recovery performance in the
Modulated Wideband Converter \cite{Eldar}.

In the first two simulations, five existing algorithms, including OMPMMV \cite{JieChen}, ReMBo
\cite{Mishali}, JLZA \cite{Hyder}, $\ell_{2,1}$ Minimization \cite{Friedlander}, and Reweighted
$\ell_{2,1}$ Minimization \cite{Reweighted}, are simulated in the same scenarios for references.
Gaussian sensing matrices are adopted in all experiments. The locations of nonzero rows in the
jointly sparse matrix are uniformly chosen at random among all possible choices, and the values of
nonzero entries are \emph{i.i.d.} standard normal distribution. We define exact recovery when the
relative error $\|{\bf X}-\hat{\bf X}\|_{\rm F}/\|{\bf X}\|_{\rm F}$ is smaller than
$1\times10^{-3}$, where $\|\cdot\|_{\rm F}$ denotes the Frobenius norm. The experiments are
conducted for $1000$ trials to calculate the probability of exact recovery or the average relative
error.

In ZAPMMV, we set $\alpha=1$, $\kappa=0.1$, $\eta=0.1$, $Q=11$, $\kappa_{\rm min}=1\times10^{-6}$, and
$T=500$. In the reference algorithms, the parameters are selected as suggested to yield the best
performance. In ReMBo algorithm, OMP is utilized for sparse recovery, and the maximum number of
iterations is $20$. In Reweighted $\ell_{2,1}$ Minimization, the reweighting procedure is conducted
for 4 iterations.

\begin{figure}[t]
\begin{center}
\includegraphics[width=3.5in]{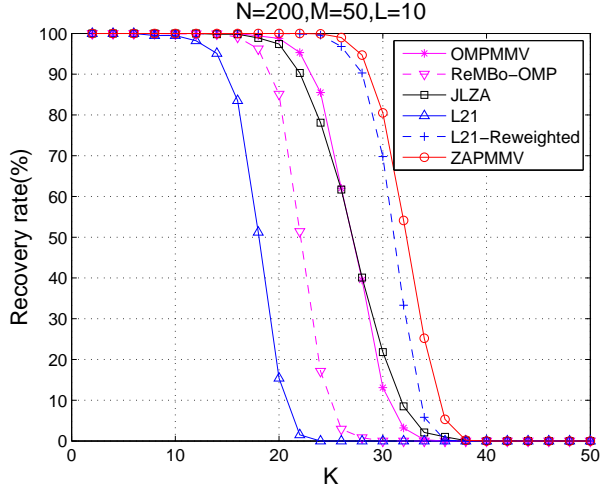}
\caption{The probability of exact recovery versus sparsity $K$.}\label{differentK}
\end{center}
\end{figure}

The first experiment tests the recovery probability with respect to the sparsity. We set $N=200$,
$M=50$, $L=10$, and the sparsity $K$ varies from $2$ to $50$. Figure~\ref{differentK} shows the
result in this scenario. As can be seen, the recovery performance of ZAPMMV exceeds those of the
other algorithms.

\begin{figure}[t]
\begin{center}
\includegraphics[width=3.5in]{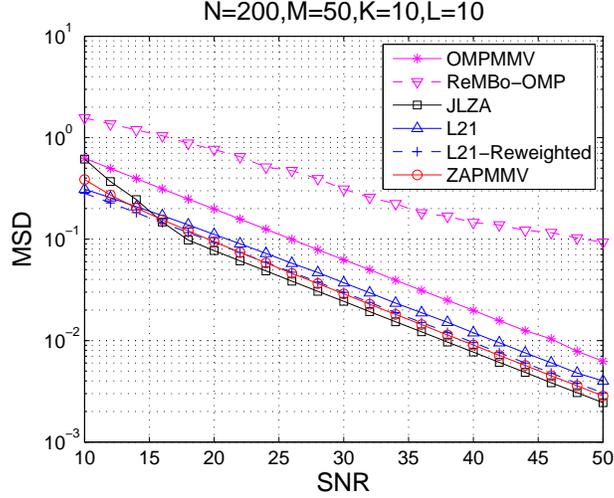}
\caption{The recovery mean squared deviation (MSD) versus SNR.}\label{MSDvsSNR}
\end{center}
\end{figure}

The second experiment studies the recovery performance against measurement noise. The measurements
are contaminated by noise as ${\bf Y}={\bf A}{\bf X}+{\bf V}$, where ${\bf V}$ denotes the zero
mean additive white Gaussian noise. Figure~\ref{MSDvsSNR} demonstrates the performance of different
algorithms in noisy environment, where the measurement signal-to-noise ratio (SNR) varies from
$10$dB to $50$dB. As can be seen from the figure, ZAPMMV is competitive in the presence of noise,
and only JLZA outperforms ZAPMMV.

The third experiment briefly compares the running time of the algorithms. Regarding the recovery
ability, only OMPMMV, JLZA, Reweighted $\ell_{2,1}$ Minimization ($\ell_{2,1}$-Reweighted), and
ZAPMMV are compared. The simulation is performed for 10 trials without perturbation of noise, and
all these algorithms can successfully recover the sparse unknowns. The average running time is
shown in Table~\ref{time}. As can be seen, OMPMMV enjoys the lowest complexity, and ZAPMMV has
lower running time than JLZA and Reweighted $\ell_{2,1}$ Minimization as the scale of the MMV
problem grows.

\begin{table}[t]
\caption{The Comparison of Average Running Time (in seconds)}
\begin{center}\label{time}
\begin{tabular}{ccccc}
\hline
  $(N,M,K,L)$ & OMPMMV & JLZA & $\ell_{2,1}$-Reweighted & ZAPMMV\\\hline
  $(1000,250,50,10)$ & $0.22$ & $8.00$ & $9.44$ & $8.04$\\
  $(2000,500,100,10)$ & $2.03$ & $50.42$ & $65.39$ & $19.03$\\
  $(3000,750,150,10)$ & $7.96$ & $169.03$ & $213.33$ & $42.19$\\
  $(4000,1000,200,10)$ & $20.04$ & $433.57$ & $457.91$ & $72.94$\\
  $(5000,1250,250,10)$ & $40.38$ & $729.52$ & $591.27$ & $106.66$\\
\hline
\end{tabular}
\end{center}
\end{table}

\begin{figure}[t]
\begin{center}
\includegraphics[width=4in]{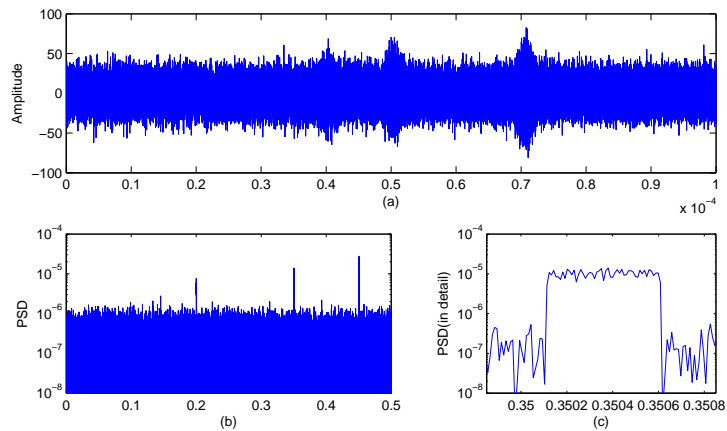}
\caption{(a) waveform and (b, c) power spectrum density (PSD) of the original signal with in-band
SNR $11$dB, $14$dB, and $17$dB, where (c) is the detail of (b). The x-axis of (a) and (b, c)
denotes time in seconds and frequency in GHz, respectively.}\label{SignalSpectrum}
\end{center}
\end{figure}

\begin{figure}[t]
\begin{center}
\includegraphics[width=3.5in]{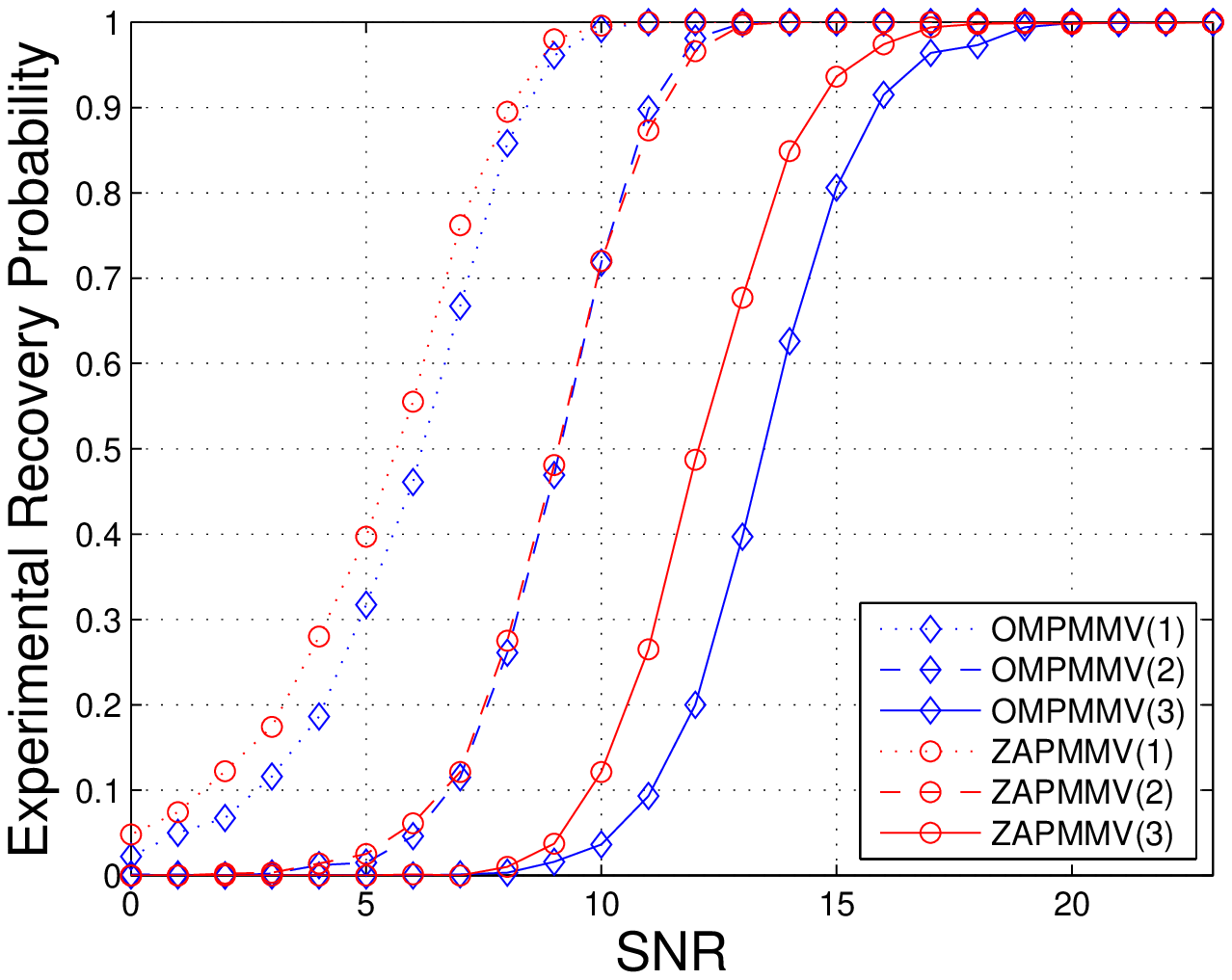}
\caption{The recovery probability of OMPMMV and ZAPMMV in the MWC simulation, where OMPMMV($i$) and
ZAPMMV($i$) denote that $i$ components are successfully detected.}\label{MWC}
\end{center}
\end{figure}

In the final experiment, OMPMMV and ZAPMMV are applied to solve the MMV problem in the MWC to
compare their recovery performance. JLZA is not included since its performance is not so fine in
this experiment. In our simulation, $30$ channels with $1$MHz ADCs are utilized to sample a sparse
signal at Nyquist rate $1$GHz, and each sampling sequence contains $999$ pulses in one period. The
original sparse signal contaminated by Gaussian white noise contains $3$ narrow-band components
(each of $0.5$MHz, and unknown location). The waveform and the power spectrum density (PSD) of the
input signals with in-band SNR $11$dB, $14$dB, and $17$dB are depicted in
Figure~\ref{SignalSpectrum}. Figure~\ref{MWC} demonstrates the probability of successfully
detecting the locations of all narrow-band components, two narrow-band components, and only one
component, where the average in-band SNR of the original signal varies from $0$dB to $23$dB. The
experiment is conducted for $1000$ trials. In this experiment, ZAPMMV recovers the signals with
higher probability than the reference method, as verified by its performance against signal noise.

\section{Conclusion}
\label{sec:conclu}

In this paper, an approximate $\ell_{2,0}$ norm is adopted to extend the zero-point attracting
projection algorithm to solve the multiple-measurement vectors problem. Computer simulations
demonstrate that the proposed algorithm outperforms several reference algorithms in respect of
sparsity, robustness to measurement noise, running time, and in the application of the Modulated
Wideband Converter. Future work may include the theoretic analysis on the convergence and recovery
accuracy of the proposed algorithm with respect to initial value and parameters, as well as
designing a ``noise-aware'' version of ZAPMMV through projecting onto the relaxed set $\{\|{\bf
Y-AX}\|_2 \le \varepsilon\}$.

\section*{Acknowledgement}

The authors appreciate the anonymous reviewers for their helpful comments to
improve the quality of this paper.


\begin{thebibliography}{00}

\bibitem{Candes}
E.~J.~Cand\`{e}s, J.~Romberg, T.~Tao, Robust uncertainty principles: Exact signal reconstruction
from highly incomplete frequency information, IEEE Transactions on Information Theory 52 (2) (2006)
489-509.

\bibitem{Donoho}
D.~L.~Donoho, Compressed sensing, IEEE Transactions on Information Theory 52 (4) (2006) 1289-1306.

\bibitem{Cotter}
S.~F.~Cotter, B.~D.~Rao, K.~Engan, K.~Kreutz-Delgado, Sparse solutions to linear inverse problems
with multiple measurement vectors, IEEE Transactions on Signal Processing 53 (7) (2005) 2477-2488.

\bibitem{Malioutov}
D.~Malioutov, M.~Cetin, A.~S.~Willsky, A sparse signal reconstruction perspective for source
localization with sensor arrays, IEEE Transactions on Signal Processing 53 (8) (2005) 3010-3022.

\bibitem{Eldar}
M.~Mishali, Y.~C.~Eldar, From theory to practice: sub-Nyquist sampling of sparse wideband analog
signals, IEEE Journal of Selected Topics in Signal Processing 4 (2) (2010) 375-391.

\bibitem{JieChen}
J.~Chen, X.~Huo, Theoretical results on sparse representations of multiple-measurement vectors,
IEEE Transactions on Signal Processing 54 (12) (2006) 4634-4643.

\bibitem{Mishali}
M.~Mishali, Y.~C.~Eldar, Reduce and boost: Recovering arbitrary sets of jointly sparse vectors,
IEEE Transactions on Signal Processing 56 (10) (2008) 4692-4702.

\bibitem{Hyder}
M.~M.~Hyder, K.~Mahata, A robust algorithm for joint-sparse recovery, IEEE Signal Processing
Letters 16 (12) (2009) 1091-1094.

\bibitem{Rakotomamonjy}
A.~Rakotomamonjy, Surveying and comparing simultaneous sparse approximation (or group-lasso)
algorithms, Signal Processing 91 (7) (2011) 1505-1526.

\bibitem{Jin}
J.~Jin, Y,~Gu, S.~Mei, A stochastic gradient approach on compressive sensing signal reconstruction
based on adaptive filtering framework, IEEE Journal of Selected Topics in Signal Processing 4 (2)
(2010) 409-420.

\bibitem{Weston}
J.~Weston, A.~Elisseeff, B.~Scholkopf, M.~Tipping, Use of the zero-norm with linear models and
kernel methods, Journal of Machine Learning Research 3 (2003) 1439-1461.

\bibitem{Friedlander}
E.~van den Berg, M.~P.~Friedlander, Probing the Pareto frontier for basis pursuit solutions, SIAM
Journal on Scientific Computing, 31 (2) (2008) 890-912.

\bibitem{Reweighted}
E.~J.~Cand\`{e}s, M.~B.~Wakin, S.~Boyd, Enhancing sparsity by reweighted $\ell_1$ minimization,
Journal of Fourier Analysis and Applications, 14 (5) (2008) 877-905.

\end{thebibliography}
\end{document}